\documentclass[prb,aps,twocolumn,superscriptaddress]{revtex4-2}

\usepackage{graphicx}
\usepackage{amsmath}
\usepackage{amssymb}
\usepackage{hyperref}
\usepackage{color}
\usepackage{lineno}
\begin{document}

	\title{Significant Chiral Magnetotransport Magnified by Multiple Weyl Nodes}

	\author{Bo~Zhang}
	\author{Junbo~Liao}
	\author{Zhentao~Huang}
	\author{Yanyan~Shangguan}
	\author{Shufan~Cheng}
	\author{Hao~Xu}
	\author{Zihang~Song}
	\author{Shuai~Dong}
	\affiliation{National Laboratory of Solid State Microstructures and Department of Physics, Nanjing University, Nanjing 210093, China}
	\author{Song~Bao}
	\email{songbao@nju.edu.cn}
	\affiliation{National Laboratory of Solid State Microstructures and Department of Physics, Nanjing University, Nanjing 210093, China}
\affiliation{Collaborative Innovation Center of Advanced Microstructures and Jiangsu Physical Science Research Center, Nanjing University, Nanjing 210093, China}
	\author{Rui~Wang}
	\email{rwang89@nju.edu.cn}
	\affiliation{National Laboratory of Solid State Microstructures and Department of Physics, Nanjing University, Nanjing 210093, China}
	\affiliation{Collaborative Innovation Center of Advanced Microstructures and Jiangsu Physical Science Research Center, Nanjing University, Nanjing 210093, China}
    \affiliation{Hefei National Laboratory, Hefei 230088, China}
	\author{Jinsheng~Wen}
	\email{jwen@nju.edu.cn}
	\affiliation{National Laboratory of Solid State Microstructures and Department of Physics, Nanjing University, Nanjing 210093, China}
	\affiliation{Collaborative Innovation Center of Advanced Microstructures and Jiangsu Physical Science Research Center, Nanjing University, Nanjing 210093, China}
	\date{\today}

	\begin{abstract}
        The intertwining of magnetism with topology is known to give rise to exotic quantum phenomena. Here, we explore the magnetotransport properties of NdAlSi, a magnetic Weyl semimetal that spontaneously breaks inversion and time-reversal symmetries and hosts a large number of Weyl nodes. We observe a significant negative magnetoresistance, which we attribute to the chiral anomaly associated with multiple Weyl nodes. Remarkably, the extracted chiral coefficient reaches approximately $52~\mathrm{m\Omega}^{-1}~\mathrm{m}^{-1}~\mathrm{T}^{-2}$, larger than many other topological materials. Additionally, we observe an exotic anomalous Hall effect with an out-of-sync behavior, where the anomalous Hall resistance does not exactly follow the field dependence of the magnetization, in contrast to that in conventional ferromagnets. These rich quantum transport phenomena, driven by the interplay between magnetism and Weyl nodes, establish NdAlSi as a prime platform for exploring the intricate topological behaviors of magnetic Weyl semimetals.
	\end{abstract}
	
	\maketitle
	
\section{Introduction}

Weyl semimetals (WSMs) host Weyl fermions as emergent quasiparticles, which behave as massless chiral particles with a nontrivial topological nature \cite{ruan2016ideal,morali2019fermi,xu2015observation}. Due to their intriguing electronic properties, such as the chiral anomaly and anomalous Hall effect (AHE), WSMs have become one of the most prominent subjects in condensed matter physics over the last decade \cite{wan2011topological,lv2015experimental,yan2017topological,armitage2018weyl,huang2015observation,huang2015weyl,neupane2014observation,liu2014stable,wang2013three,PhysRevLett.117.236401,manna2018heusler}. Recently, the $R$Al$X$ family ($R=$~Lanthanide and $X=$~Si/Ge) has emerged as a new class of magnetic Weyl semimetals. These materials have gathered significant attention due to their unique electronic properties driven by the simultaneous breaking of inversion and time-reversal symmetries \cite{gaudet2021weyl,PhysRevB.108.205143,PhysRevB.108.L161106,PhysRevB.109.L201108,PhysRevX.14.021012}. Among these materials, NdAlSi stands out due to its distinguished features, particularly its helical magnetism, which is mediated by the Weyl topological electronic structure \cite{gaudet2021weyl}.

\begin{figure}[t]
	\centerline{\includegraphics[width=9.cm]{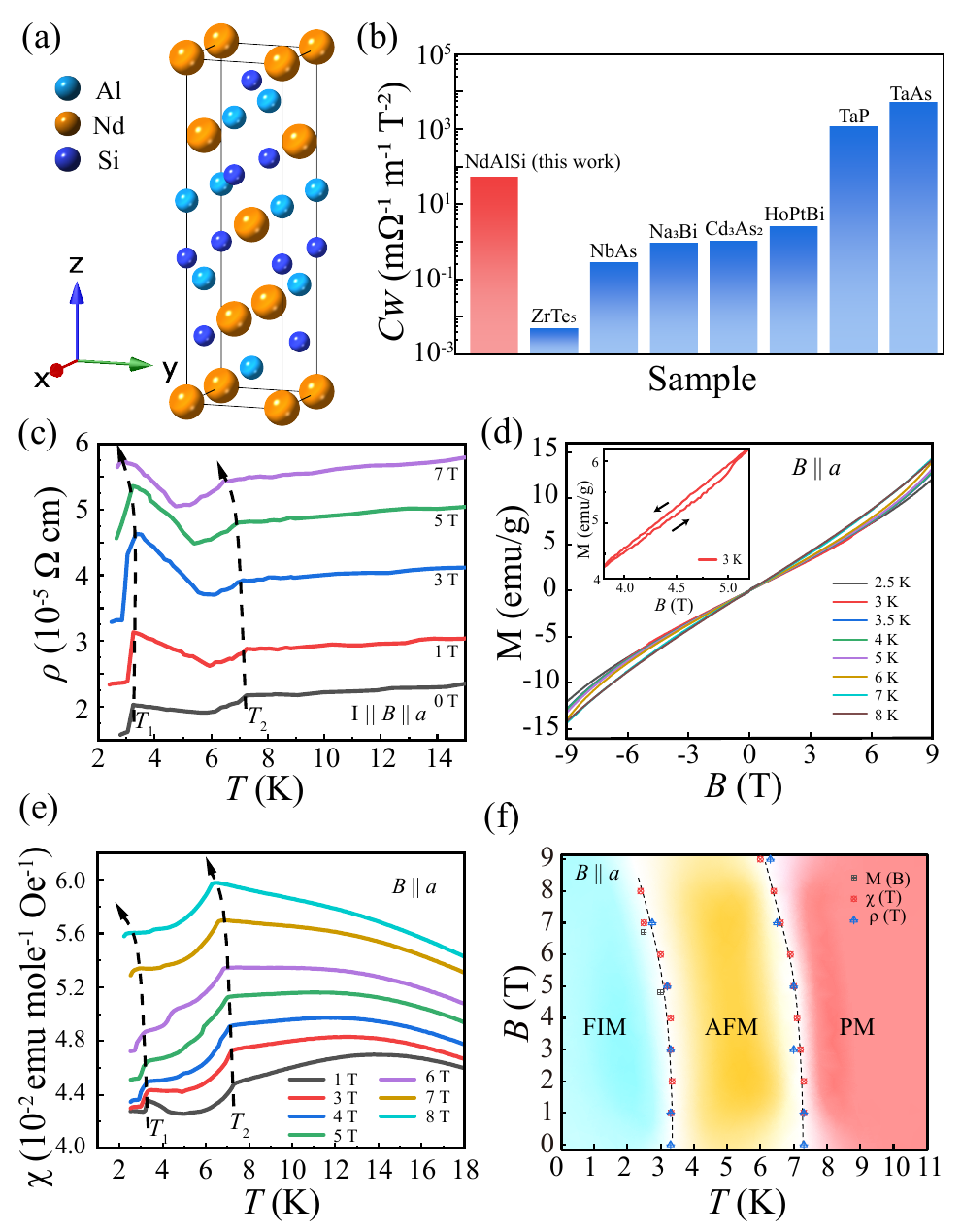}}
	\caption{\label{fig1}
	(a) Schematic crystal structure of NdAlSi with a non-centrosymmetric space group, $I4_{1}md$~(No.~109). 
	(b) Chiral coefficient ($C_{\rm W}$) of various topological metals extracted at low temperatures 
	\cite{li2016chiral,li2017negative,xiong2015evidence,li2016negative,10.1063/5.0026956,arnold2016negative,zhang2016signatures}. 
	(c) Temperature dependence of resistance under different in-plane magnetic fields. The current is parallel to the fields, whose direction is along the crystalline $a$-axis. The two dashed arrows mark two critical temperatures, $T_{1}$ and $T_{2}$. The data are separated by a vertical shift of $5 \times 10^{-6}~\Omega~\text{cm}$.
	(d) Field dependence of magnetization at different temperatures. The external field lies in the $a$-$b$ plane. The inset shows the enlarged image, from which a hysteresis is observed at high fields and low temperatures.
	 (e) Temperature dependence of the susceptibility at different in-plane fields. The external field is along the $a$-axis. The two dashed arrows mark two critical temperatures, $T_{1}$ and $T_{2}$. The curves are vertically offset by $1.5 \times 10^{-3}~\text{emu mole}^{-1}~\text{Oe}^{-1}$ for clarity.
	(f) Schematic phase diagram consisting of ferrimagnetic (FIM), antiferromagnetic (AFM), and paramagnetic (PM) phases for field applied along the $a$-axis.}
	
\end{figure}

NdAlSi possesses a non-centrosymmetric lattice structure [Fig.\ref{fig1}(a)], allowing Weyl nodes to emerge even in the paramagnetic (PM) state due to the breaking of inversion symmetry. Previous band structure calculations have revealed that NdAlSi hosts 40 Weyl nodes throughout the entire Brillouin zone \cite{gaudet2021weyl}. Furthermore, it exhibits several distinct magnetic phases at relatively low temperatures. In these magnetic ordering phases, time-reversal symmetry is spontaneously broken, resulting in band splitting and a consequent increase in the number of Weyl points. Specifically, the number of Weyl points increases to 52, 56, and 56 in the ferrimagnetic (FIM), antiferromagnetic (AFM), and spin-polarized (SP) states, respectively \cite{li2023emergence,gaudet2021weyl}. This topological system, with its various magnetic phases, offers an unprecedented opportunity to explore the rich topological physics, which is associated with magnetism, spin textures, and electronic transport \cite{burkov2014anomalous}. Several intriguing questions naturally arise: How does the large number of Weyl nodes affect electronic transport? Can topological response be enriched by magnetism? Can unconventional magnetotransport properties emerge in this material?

In this work, we conduct comprehensive magnetotransport measurements on NdAlSi and discover magnetism-enriched topological transport behaviors that are absent in other WSMs. Through temperature- and angle-dependent measurements under in-plane magnetic fields, we observe a significant negative magnetoresistance in this material. The chiral coefficient ${C}_{\rm W}$, which quantifies the strength of the chiral anomaly \cite{son2013chiral,niemann2017chiral,son2013chiral,li2017negative,huang2015observation,burkov2015negative,arnold2016negative,wang2016gate}, is found to be as high as $52~\text{m}\Omega^{-1}~\text{m}^{-1}~\text{T}^{-2}$, larger than many other topological materials, as summarized in Fig.~\ref{fig1}(b). This pronounced manifestation of the chiral anomaly is naturally attributed to the large number of Weyl nodes near the Fermi surface, which act as monopoles of Berry curvature. In addition to the large negative magnetoresistance, we also observe an unconventional AHE under perpendicular magnetic fields. Unlike in ferromagnets, the anomalous Hall resistance in NdAlSi is not simply ''proportional'' to the magnetization but exhibits an out-of-sync field dependence with the $M$-$B$ curve. Furthermore, the anomalous Hall component, $R_{A}$, is nearly four orders of magnitude larger than the ordinary Hall component $R_H$. This observation is consistent with the large chiral coefficient ${C}_{\rm W}$, as both phenomena stem from the substantial Berry curvature induced by the multiple Weyl nodes. Lastly, we detect quantum oscillations in the magnetotransport data, further highlighting the intricate interplay between magnetism and topology. Our results demonstrate that magnetism can indeed enhance the topological electromagnetic response of WSMs, leading to fascinating transport phenomena driven by the intertwining of magnetism and topology.

\section{Experimental Details}

High-quality single crystals of NdAlSi were grown using the self-flux method with aluminum as the flux, as described in Refs.~\onlinecite{gaudet2021weyl,wang2022ndalsi}. The starting materials, Nd, Al, and Si, were weighed in a molar ratio of $n{\rm(Nd)}:n{\rm(Al)}:n{\rm(Si)} = 1:10:1$. These materials were ground, mixed to ensure homogeneity, and then loaded into an alumina crucible. The crucible containing the mixture was placed inside a quartz ampule, which was sealed under high vacuum. The ampule was heated to 1150~$^{\circ}$C and maintained at this temperature for 12 hours. It was then cooled to 720~$^{\circ}$C at a controlled rate of 0.05~$^{\circ}$C/min and centrifuged to remove the excess flux. The resulting samples were plate-like with well-defined square edges.

To investigate the transport properties of NdAlSi, angle-dependent magnetoresistance measurements were performed using a standard four-probe configuration in a Quantum Design Physical Property Measurement System (PPMS-9T). The electrical current was applied along the $a$-axis, and the magnetic field was applied within the $a$-$b$ plane. This setup allows for a detailed examination of the magnetoresistance behavior relative to the magnetic field orientation with respect to the crystal lattice, as shown in the inset of Fig.~\ref{fig2}(c). Hall resistance measurements were conducted with the magnetic field applied along the $c$-axis and the current flowing along the $a$-axis. The Hall resistance data, $\rho_{xy}$, were symmetrized using the function $\rho_{xy} = [\rho_{xy}(B) + \rho_{xy}(-B)]/2$. Magnetization measurements were performed with the PPMS equipped with a vibrational superconducting quantum interference device magnetometer kit.

\section{Results and discussions}

\subsection{Phase diagram analyzed from resistivity and in-plane magnetization}

We first present the measured in-plane resistivity in Fig.~\ref{fig1}(c), where the applied magnetic field is parallel to the current. At zero field, the resistivity shows a pronounced peak at $T_{1}= 3.3$~K and a kink at $T_{2} = 7.2$~K. According to a previous neutron diffraction study \cite{gaudet2021weyl}, these temperatures correspond to two magnetic phase transitions. Specifically, at low temperatures ($T < T_1$), the material is in a commensurate FIM phase, which transitions to an incommensurate AFM phase for $T_{1} < T < T_{2}$. Above $T_{2}$, the material enters a PM phase. Notably, there is an unusual resistivity minimum around 6~K, which has been attributed to several possible factors \cite{suzuki2019singular,gaudet2021weyl}, including variations in spin excitations \cite{PhysRevB.9.272}, changes in the Fermi surface \cite{PhysRevB.99.035142}, and domain wall scattering \cite{PhysRevLett.79.5110}. Determining the exact origin of this minimum will require further investigation. Applying an external magnetic field within the $a$-$b$ plane increases the resistivity while preserving the overall behavior, with only slight shifts in the critical temperatures $T_{1}$ and $T_{2}$, as indicated by the dashed curves in Fig.~\ref{fig1}(c).

The in-plane magnetization as a function of field and temperature is shown in Fig.~\ref{fig1}(d) and (e), respectively. Figure~\ref{fig1}(e) exhibits similar features to the resistance behavior in Fig.~\ref{fig1}(c), with two critical temperatures clearly identified. The field-dependent magnetization in Fig.~\ref{fig1}(d) varies smoothly with $B$, while hysteresis is observed at low temperatures, such as at 3~K in the field range of 4~T to 5~T, as shown in the inset of Fig.~\ref{fig1}(d). This behavior indicates the presence of magnetic domains between FIM and AFM phases within the material, which may also contribute to the hysteresis observed in the magnetoresistance, as discussed in the following section.

By incorporating our new resistivity and magnetization data alongside previously reported results, we construct a phase diagram with field applied along the $a$-axis, shown in Fig.~\ref{fig1}(f). A phase diagram with field applied along the $c$-axis is provided in Fig.~\ref{fig5}(b) in Appendix A. These phase diagrams provide a foundation for understanding the magnetotransport phenomena discussed in the following sections.

\begin{figure}[tb]
    \centerline{\includegraphics[width=9.cm]{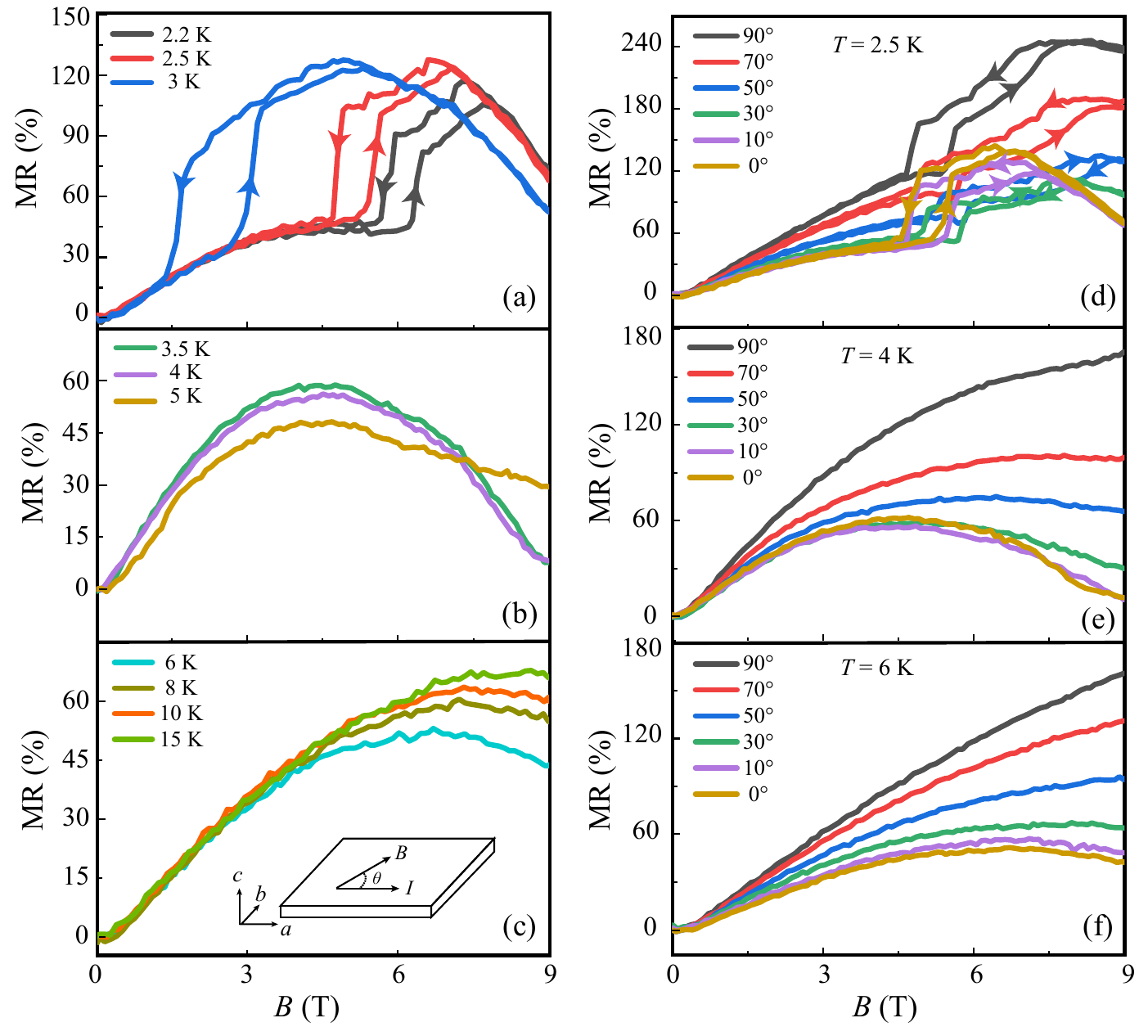}}
    \caption{\label{fig2}
    Magnetoresistance and its temperature and angle dependence. The magnetoresistance ($MR$) is calculated as ${\rm MR}(\%)=\frac{\rho (B)-\rho (0)}{\rho (0)} \times 100\%$, where $\rho (B)$ represents the resistivity under a magnetic field, and $\rho (0)$ denotes the resistivity at zero field. (a), (b), and (c) present the temperature dependence of the MR. (d), (e), and (f) illustrate the angle dependence of the magnetic resistance at temperatures of 2.5, 4, and 6~K, respectively. The inset in (c) is a schematic plot of the measurement setup.
    }
\end{figure}

\subsection{Negative magnetoresistance}
The in-plane field direction [{\it i.e.}, the angle $\theta$ in the inset of Fig.~\ref{fig2}(c)] and the temperature serve as two effective parameters for studying the magnetotransport properties. We first fix $\theta$ to be zero to explore the magnetoresistance as a function of temperature, as shown in Fig.~\ref{fig2}(a)-(c). 

Fig.~\ref{fig2}(a) displays the magnetoresistivity for $T < T_{1}$, where the FIM state is stabilized. At $T = 2.2$~K, the magnetoresistivity shows a monotonic increase followed by a sharp jump and hysteresis around 6~T. The maximum resistivity occurs around 7~T, after which the slope becomes negative at higher fields. As the temperature increases, both the jump and hysteresis shift to lower fields, with the maximum resistivity moving to 4.5~T at $T = 3$~K.  The sharp jump in the resistivity at low temperatures, as shown in Fig.~\ref{fig2}(a), corresponds to the magnetic phase transition from the FIM to the AFM phase, as indicated by the phase boundary in Fig.~\ref{fig1}(f). Since magnetization also shows hysteresis at low temperatures [{\it i.e.}, 3~K data in the inset of Fig.~\ref{fig1}(d)], it suggests that NdAlSi contains magnetic domains between these two magnetic phases, each with its own Fermi surface. The mismatch at the boundaries of these domains leads to electron scattering, which results in high resistance and hysteresis \cite{suzuki2019singular}. As the temperature rises, the shift of hysteresis to lower fields and its eventual disappearance is consistent with the in-plane phase diagram [Fig.~\ref{fig1}(f)] and a reduction in the energy needed to unify the magnetic domains. A similar mechanism is observed in its sister compound CeAlGe \cite{suzuki2019singular}.

For $3.5$~K $\lesssim T \lesssim$ 5~K, corresponding to the AFM state, the behavior changes as shown in Fig.~\ref{fig2}(b). Both the jump and hysteresis are absent, and the maximum magnetoresistance remains around 4.5~T regardless of temperature. Negative magnetoresistance is clearly observed at higher fields. In the PM phase ($T > 6$~K), the maximum magnetoresistance and the region of negative magnetoresistance shift to higher fields with increasing temperature, as shown in Fig.~\ref{fig2}(c).

We further investigate angle-dependent magnetoresistance by varying the direction of the external field from $\theta = 0^\circ$ ($i.e.$, $B \parallel I$) to $\theta = 90^\circ$ ($i.e.$, $B \perp I$). The magnetoresistance for three representative temperatures, corresponding to FIM, AFM, and PM phases, is shown in Fig.~\ref{fig2}(d)-(f). At 2.5~K, the magnetoresistance decreases as the angle is reduced, exhibiting hysteresis around 4.5~T. Negative magnetoresistance is observed for small angles ($\theta < 10^\circ$) and relatively large fields (exceeding 7~T). At 4~K, the magnetoresistance shows significant angle dependence. For $\theta = 90^\circ$ ($i.e.$, the magnetic field perpendicular to the current), no negative magnetoresistance is observed. As $\theta$ decreases, the magnetoresistance curves bend downward, showing a large region with a negative slope. At 6~K, the magnetoresistance continues to exhibit angle dependence similar to that at 4~K. Although negative magnetoresistance is observed for $\theta < 30^\circ$, its potential occurrence at larger angles and higher fields beyond the field range in our experiments, cannot be excluded.

Consistent with previous reports on other topological semimetals \cite{lv2015experimental,yang2015weyl,xu2015experimental,yan2017topological,LI2018535}, negative magnetoresistance is attributed to a nonzero Berry curvature and approximately conserved chiral charge density. The Weyl nodes act as monopoles of Berry curvature in the momentum space \cite{zhang2016signatures}, driving unique topological electromagnetic responses described by the chiral anomaly. Specifically, under a nonzero $\mathbf{B} \cdot \mathbf{E}$ field, a charge pumping effect occurs between Weyl nodes, causing an imbalance in the quasiparticle population of Weyl cones with opposite chiralities \cite{son2013chiral,burkov2014chiral}, which results in negative magnetoresistance.

\begin{figure}[tb]
    \centerline{\includegraphics[width=9.cm]{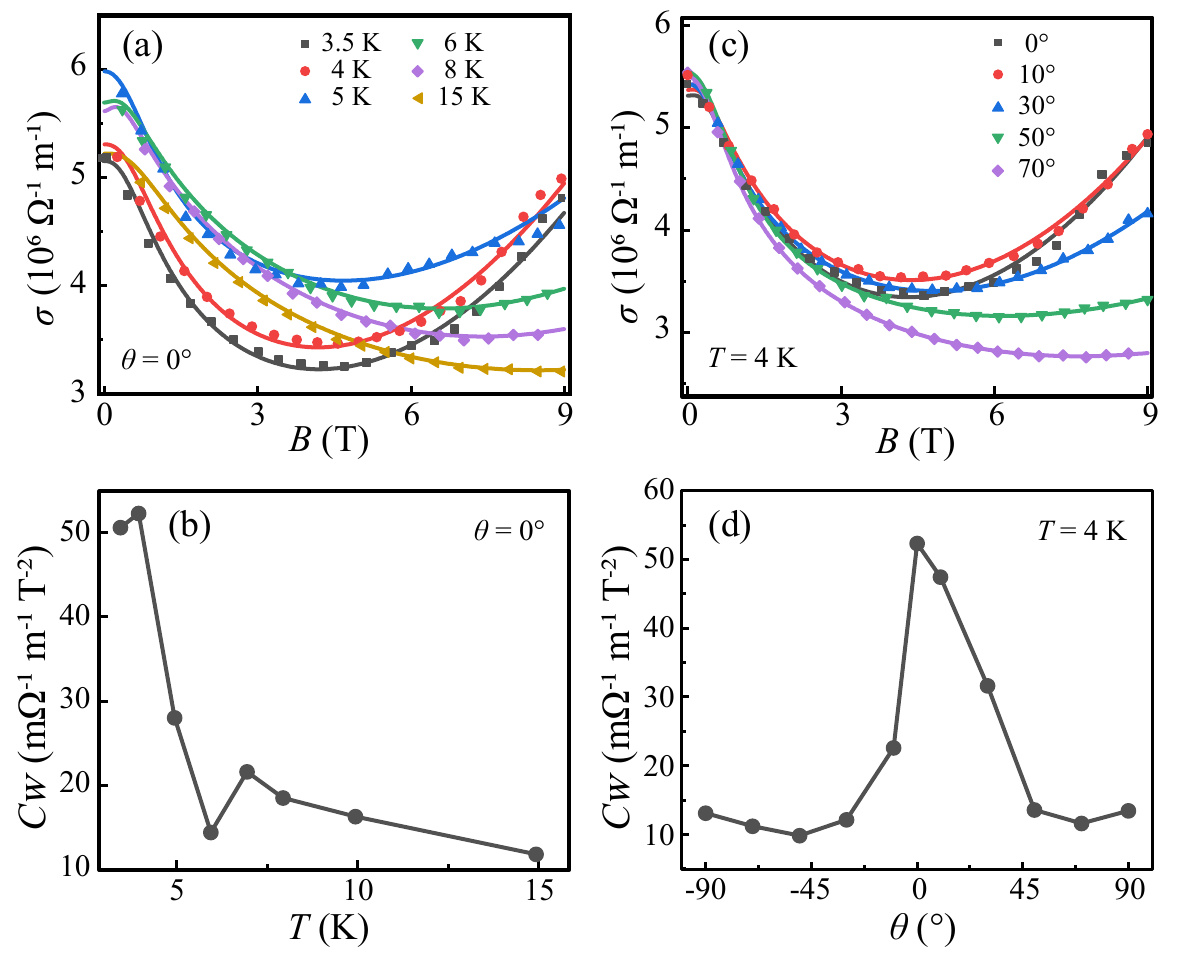}}
    \caption{\label{fig3}
    Analyses of the magnetoconductivity. Magnetoconductivity as a function of temperature (a) and angle $\theta$ (c). Experimental data are shown by the markers, while the solid curves are the fittings according to Eq.~\eqref{eq1}. (b) and (d) are the chiral coefficients ($C_{\rm W}$) extracted from the fittings in (a) and (c), respectively.
    }
\end{figure}

To analyze the negative magnetoresistance, it is useful to examine the magnetoconductance. The measured magnetoconductance is shown in Fig.~\ref{fig3} for varying temperatures (a) and angles (b). In a Weyl semimetal, the magnetoconductance can be described by a semiclassical formula that accounts for contributions from Weyl nodes and their Berry curvatures \cite{armitage2018weyl,son2013chiral,burkov2015chiral}:

\begin{equation}
	\label{eq1}
	\sigma(B) =   C_{\mathrm{W}} B^{2} - C_{\mathrm{WAL}} \left( \frac{\sqrt{B} B^{2}}{B^{2} + B_{c}^{2}} + \frac{\gamma B^{2} B_{c}^{2}}{B^{2} + B_{c}^{2}}\right) 
	 +\sigma_{0}.
\end{equation}

The first term, $C_{\rm W} B^{2}$, represents the effect of the chiral anomaly, an intrinsic contribution from the Berry curvature associated with the Weyl nodes. $C_{\rm W}$ is the chiral coefficient, given by $N \cdot e^{4}\tau_{a}/[4\pi^{4}\hbar^{4}g(E_{\rm F})]$ \cite{son2013chiral,burkov2015chiral}, where $N$, $e$, $\tau_{a}$, and $g(E_{\rm F})$ are the number of Weyl node pairs, electron charge, the internode relaxation time,  and the density of states at the Fermi level, respectively.

The second term, proportional to $C_{\mathrm{WAL}}$, accounts for the three-dimensional weak anti-localization (WAL) effect related to the Weyl cones. The WAL effect is known to exhibit a $-B^{2}$ dependence at small fields and a $-\sqrt{B}$ dependence at higher fields, with a crossover at a critical field, $B_{c}$. This term explains the initial steep rise in magnetoresistance at small fields (Fig.~\ref{fig2}). The last term, $\sigma_{0}$, represents the positive longitudinal magnetoresistance due to conventional Drude conductivity, which also appears in other Weyl semimetals like TaAs \cite{lv2015experimental,huang2015observation,yang2015weyl}. For parallel magnetic fields, the Drude conductivity remains constant \cite{smith2001classical}.

As shown by the solid curves in Figs.~\ref{fig3}(a) and \ref{fig3}(c), the experimental data fit well with Eq.~\eqref{eq1}, from which we extract the chiral coefficient $C_{\mathrm{W}}$. Figs.~\ref{fig3}(b) and \ref{fig3}(d) display $C_{\mathrm{W}}$ as a function of temperature and angle, respectively. $C_{\mathrm{W}}$ reaches its maximum value when the electric and magnetic fields are parallel, consistent with the chiral anomaly mechanism associated with Weyl nodes.

We also measured the magnetoresistance with the current parallel to [110], the modulation vector associated with the magnetic order. As shown in Appendix C and Fig.~\ref{fig7}, the negative magnetoresistance is less pronounced in the $B \parallel I \parallel [110]$ configuration compared to $B \parallel I \parallel a$, suggesting that the nesting vector along [110] does not enhance the negative magnetoresistance. The most evident negative magnetoresistance effect is observed when the field is aligned along the [100] direction.

In real materials, the origin of negative magnetoresistance can be diverse, such as from current jetting or the suppression of spin scattering. However, as discussed in Ref.~\onlinecite{PhysRevB.108.L161106}, heat transport measurements in NdAlSi, which are robust against current distribution effects, confirm the intrinsic nature of its negative magnetoresistance, ruling out current jetting as the primary cause. Additionally, our angular dependence measurements in Fig.~\ref{fig8} show no signatures of current jetting, such as dips, humps, or negative voltage. While spin scattering suppression could contribute to negative magnetoresistance in some materials, the unique topological properties of NdAlSi and its Weyl nodes strongly suggest that the observed negative magnetoresistance is driven by the chiral anomaly.

This interpretation is further supported by the extracted $C_{\mathrm{W}}$, which exhibits temperature and angle dependence consistent with the charge-pumping mechanism between Weyl nodes. Moreover, the intrinsic chiral anomaly is characterized by a quadratic dependence of electrical conductivity on the magnetic field ($B^2$), a behavior observed in our high-field measurements shown in Fig.~\ref{fig3}(a) and (c).The relatively large chiral coefficient in NdAlSi reflects the substantial number of Weyl points $N$ inherent to this material \cite{li2023emergence,gaudet2021weyl}.

\begin{figure}[tb]
    \centerline{\includegraphics[width=9.cm]{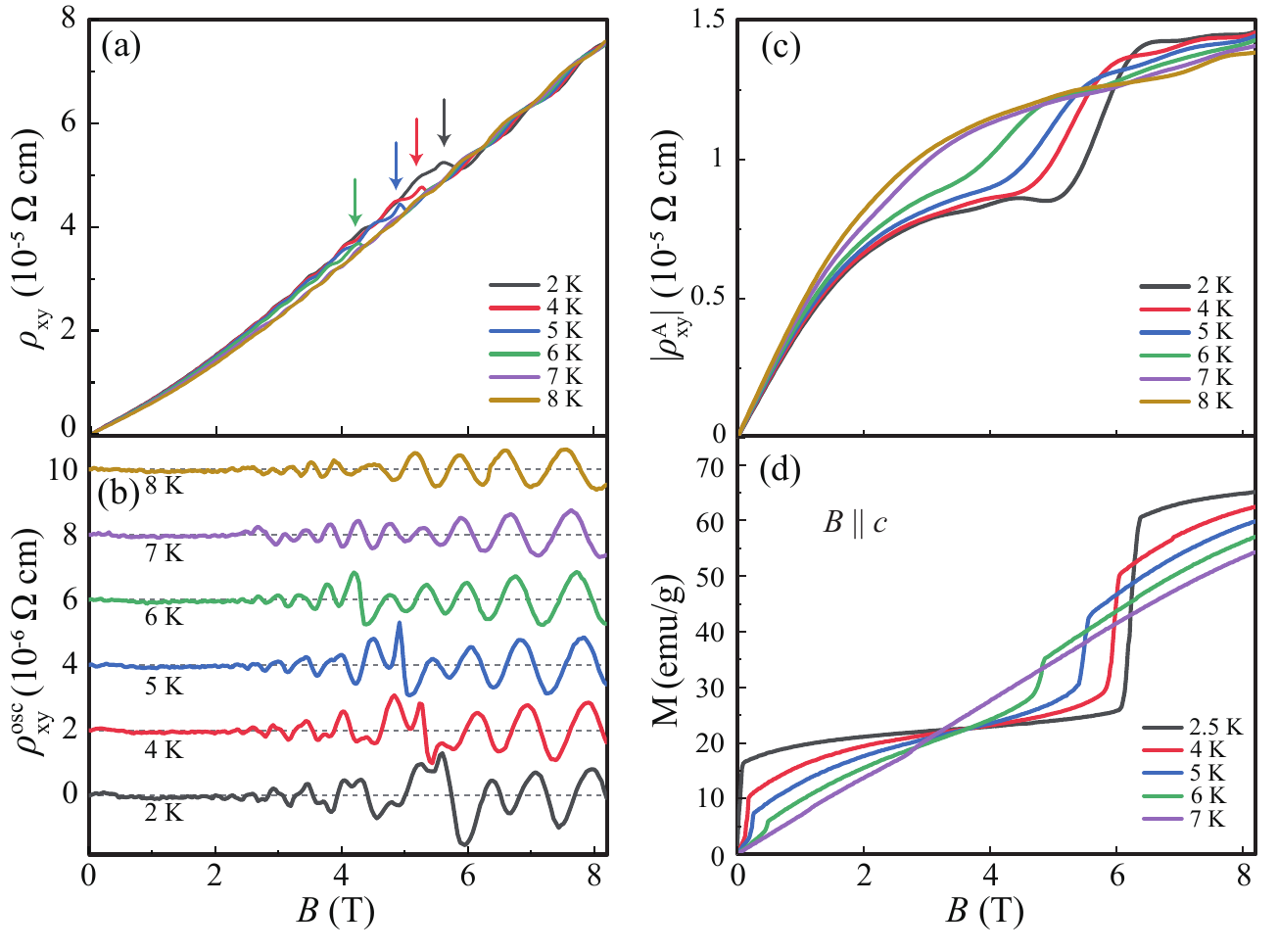}}
    \caption{\label{fig4}
    Analyses of the anomalous Hall effect. (a) The Hall resistivity ($\rho_{xy}$) as a function of magnetic field along the easy axis ($B \parallel c$). Arrows show the positions of the kinks. (b) The oscillation part of $\rho_{xy}$ at each temperature. The data are separated by a vertical shift of $2 \times 10^{-6}~\Omega~\text{cm}$. (c) Anomalous Hall resistivity ($\rho_{xy}^{A}$) versus magnetic field, extracted from the relation discussed in the main text. (d) Magnetization as a function of magnetic field ($B$) along the magnetic easy axis ($B \parallel c$).
    }
\end{figure}

\subsection{Anomalous Hall Effect}

In magnetic Weyl semimetals that break time-reversal symmetry, a non-zero Berry curvature leads to an AHE. Unlike the ordinary Hall effect, the AHE depends not only on the external magnetic field but also on the intrinsic magnetism of the material. The in-plane magnetization in Fig.~\ref{fig1}(d) evolves smoothly with field, indicating no field-induced phase transitions. In contrast, the out-of-plane magnetization, shown in Fig.~\ref{fig4}(d), exhibits step-like plateaus at each temperature, suggesting the presence of different magnetic phases. To understand how these magnetic phases influence the AHE, we focus on magnetotransport properties with perpendicular fields along the easy axis, {\it i.e.}, the $c$-axis.

Figure~\ref{fig4}(a) shows that the Hall resistance deviates from linearity, indicating the presence of anomalous contributions in addition to the conventional Hall effect. The sign of the Hall resistivity, $\rho _{xy}$, is positive, indicating hole-like carriers. This observation is consistent with previous reports \cite{PhysRevMaterials.7.034202,wang2022ndalsi}. A notable kink around 6.2~T at 2~K shifts to lower fields with increasing temperature and disappears around 7~K. As highlighted in Fig.~\ref{fig5}(b) (Appendix A), this kink coincides with the phase boundary between the FIM and SP states. Thus, the kinks in the Hall resistance are attributed to the FIM-SP transition and the corresponding change in the Fermi surface. Additionally, periodic oscillations in the Hall resistance at high fields suggest Shubnikov-de Haas quantum oscillations, resulting from the Landau levels crossing the Fermi energy.

To analyze these results, we decompose the Hall resistance into three components: $\rho_{xy} = R_{0}B + \rho_{xy}^{\rm A} + \rho_{xy}^{\rm osc}$ (Refs.~\onlinecite{pugh1930hall,pugh1932hall}). Here, $R_{0}B$ represents the conventional Hall resistance, with the coefficient $R_{0}$ extracted by fitting the data in the high-field range of [6, 9] T. $\rho_{xy}^{\rm A}$ is the anomalous Hall resistance, and $\rho_{xy}^{\rm osc}$ accounts for quantum oscillations. Fitting the experimental data to this model at different temperatures allows extraction of each component. At $T = 2$~K, $R_{0} = 1.09129 \times 10^{-5}~\Omega~\text{cm}~\text{T}^{-1}$, indicating a low electron concentration ($n_{\rm e} \sim 10^{20}~\text{cm}^{-3}$, or approximately 0.0003 electrons per unit cell), consistent with the results from CeAlSi~(Ref.~\onlinecite{yang2021noncollinear}) and the low density of states at the Fermi level. Fig.~\ref{fig4}(b) shows the extracted $\rho_{xy}^{\rm osc}$, revealing clear Shubnikov-de Haas oscillations at high fields. The observed kinks align with the FIM-SP transition, with additional details provided in Fig.~\ref{fig6} (Appendix B).

However, unlike NdAlGe \cite{PhysRevMaterials.7.034202}, which exhibits a similar magnetic structure and magnetization plateau, the $\rho_{xy}^{\rm A}$-$B$ curve of NdAlSi does not perfectly align with the jump in the $M$-$B$ curve near zero field. 

There are several possible mechanisms for this out-of-sync behavior of AHE. Cong \emph{et\,al.}\cite{li2024nonhermitianboundarysurfaceselective} proposed that non-Hermitian effects arising from surface reconstruction on the Al-terminated surface of NdAlSi could be responsible for its unique AHE behavior. This reconstruction introduces non-Hermitian terms into the effective Hamiltonian of the surface states, thereby causing the out-of-sync behavior, as explained below.

Theoretically, the intrinsic Berry curvature contribution to the Hall conductance is expressed as:
\begin{equation}\label{eq2}
	\sigma_{ij} = \epsilon_{ijk} \frac{e^2}{\hbar} \sum_n \int \frac{d\mathbf{k}}{(2\pi)^3} n_F(\epsilon_n(\mathbf{k})) b_n(\mathbf{k}),
\end{equation}
where $b_{n}(\mathbf{k}) = i \nabla_{\mathbf{k}} \times \left\langle n, \mathbf{k} \left| \nabla_{\mathbf{k}} \right| n, \mathbf{k} \right\rangle$ is the Berry curvature of electrons in the $n$-th band, and $n_F$ is the Fermi-Dirac distribution. For magnetic Weyl semimetals studied here, $b_{n}(\mathbf{k})$ is related to and proportional to the magnetization. 

The effective non-Hermitian effect arising from surface state reconstruction reduces the scattering time $\tau$ of electrons, either due to the additional scattering channel into non-topological surface states or due to dissipation caused by the non-Hermitian boundary \cite{li2024nonhermitianboundarysurfaceselective}. According to Kohler's rule, the change in magnetoresistance is determined by its scaling with respect to $B\tau$. Consequently, a reduced $\tau$ at a given magnetic field $B$ would have a similar effect as reducing the magnetic field at fixed $\tau$. Physically, a smaller $\tau$ implies that electrons are scattered before fully experiencing the effect of the magnetic field (e.g., completing cyclotron motion). Hence, the transport behavior can be understood by effectively reducing the magnetic fields. Therefore, the surface reconstruction would effectively reduce $B$, which significantly delays the magnetization and thus the development of $b_{n} (k)$. This mechanism qualitatively explains the 
observed out-of-sync anomalous Hall behavior in NdAlSi.

Besides, the presence of magnetic domain walls introducing additional complexities could also account for the out-of-sync behavior. Each domain exhibits its own Fermi surface with different orientations of Weyl node pairs. While each domain satisfies $\sigma_{xy}^{\rm A} = {e^2 \kappa}/{4\pi^2}$, where $\kappa$ is the distance between a pair of Weyl nodes, scattering off domain walls leads to high longitudinal resistance $\rho_{xx}$ and a small $\rho_{xy}^{\rm A}$ at zero field. Under magnetic fields, although magnetic domain walls align with the net magnetization along the $c$-axis [Fig.~\ref{fig4}(d)], the alignment of Fermi surfaces with the magnetic field lags behind, making $\rho_{xy}^{\rm A}$ prominent only at higher fields.

Other possibilities are also considered. For example, regarding the multi-carrier scenario, where the non-$B$-linear Hall effect could arise, we explore the two-band Drude model fitting in the FIM state within the field range of 0-5~T ~(Ref.~\onlinecite{PhysRevB.101.125119}). However, this approach does not successfully fit both $\rho_{xy}$ and $\rho_{xx}$ simultaneously. In contrast, an earlier study \cite{wang2022ndalsi} successfully applied a single-band fitting of Hall resistivity data around zero field (within 0.5~T), yielding a carrier density on the order of $10^{20}~{\rm cm}^{-3}$ and a mobility of $10^{3}~{\rm cm}^{2}/({\rm V \cdot s})$. Thus, the multi-carrier scenario is unlikely in this case. Additionally, since the non-linear Hall resistance is observed in magnetic field regions where the material remains in the same magnetic phase ({\it i.e.} 2 K, 0-4~T), changes in the magnetic field respect the magnetic structure, and the electronic structure remains nearly intact. Therefore, potential changes in the electronic structure as a mechanism for the non-$B$-linear Hall effect can also be excluded.

Despite the out-of-sync behavior differences near zero field, both $\rho_{xy}^{\rm A}$ and $M$ exhibit a step-like jump around 5-6~T at low temperatures, which is gradually smeared out with increasing temperature. This jump in $\rho_{xy}^{\rm A}$ reflects the kink observed in Fig.~\ref{fig4}(a). This behavior is linked to the transition between the FIM phase with an up-up-down spin configuration ($\uparrow\uparrow\downarrow$) and the fully polarized state ($\uparrow\uparrow\uparrow$) (Appendix A). Therefore, although $\rho_{xy}^{\rm A}$ and $M$ show differences, the AHE is closely related to magnetization and characterizes time-reversal symmetry breaking.

To quantitatively describe the contributions of the ordinary and anomalous Hall effects, we use the relation $\rho_{xy}/{B} = R_{0} + R_{\rm A} {M}/{B}$~(Ref.~\onlinecite{burkov2014anomalous}), where $R_{\rm A}$ represents the AHE contribution. The fitting results show that the anomalous component, $R_{\rm A}$, is nearly four orders of magnitude larger than the ordinary component, underscoring the dominance of AHE in transport.

The anomalous Hall conductivity, $\sigma_{xy}^{\rm A}= - {\rho_{xy}^{A}}/({\rho_{xx}^{2} + \rho_{xy}^{2}})$, reaches an exceptionally large value of $947~\Omega^{-1}~\text{cm}^{-1}$ at 2~K and 9~T. This large value is consistent with the expected magnitude of intrinsic Berry curvature contributions, typically in the order of $\sigma_{xy}^{\rm A} = {e^{2}}/{ha} \sim 10^{2}~\Omega^{-1}~\text{cm}^{-1}$~(Refs.~\onlinecite{manna2018heusler, xiao2010berry}), where $h$ and $a$ are the Planck constant and average lattice constant, respectively. This suggests that the intrinsic Berry curvature plays a dominant role in the AHE at high field.

In magnetic semimetals, theoretical work indicates that extrinsic mechanisms such as skew scattering and side jump are negligible when the Fermi energy is close to the Weyl nodes, as their contributions cancel out after averaging over the Fermi surface \cite{burkov2014anomalous}. We compute the longitudinal electrical conductivity, $\sigma_{xx}$, to be $\sim 5.707 \times 10^{3}~\Omega^{-1}~\text{cm}^{-1}$, indicating that our sample is in the dirty limit typical of metallic materials due to scattering by impurities or domain walls. Extrinsic contributions, particularly skew scattering, become significant only in the ultra-clean limit. Therefore, extrinsic mechanisms are largely excluded, consistent with Ref.~\onlinecite{burkov2014anomalous}.

Our data suggest that the observed AHE arises primarily from intrinsic Berry curvature associated with multiple pairs of Weyl nodes in NdAlSi~(Ref.~\onlinecite{lyu2020nonsaturating}). The substantial Berry curvature is a dominant factor, surpassing extrinsic skew scattering and side jump contributions. Furthermore, the intrinsic AHE shows a distinct behavior compared to conventional ferromagnets, likely due to the non-Hermitian effect and the interplay between Berry curvature and domain walls.

\section{Conclusions}

In summary, we have systematically investigated the negative magnetoresistance and AHE in the magnetic Weyl semimetal NdAlSi. The large number of Weyl nodes in this material plays a crucial role, leading to an exceptionally large chiral coefficient that rivals the highest value reported and a significant anomalous Hall conductance that is nearly four orders of magnitude greater than the ordinary Hall component. These transport phenomena are further influenced by the presence of distinct magnetic phases and domain walls.
Our findings reveal a novel anomalous Hall resistance with a field dependence that deviates from the magnetization behavior, contrasting sharply with conventional ferromagnets. This discrepancy highlights the non-Hermitian effect in Weyl material NdAlSi. The comprehensive magnetotransport data underscore the presence of multiple Weyl nodes and demonstrate how their interactions with magnetic domains produce complex transport behaviors.
Overall, NdAlSi emerges as an ideal material platform for studying the interplay of Weyl nodes, Berry curvature, and magnetic effects in magnetic Weyl semimetals.

\begin{acknowledgments}
The work was supported by National Key Projects for Research and Development of China with Grant Nos.~2021YFA1400400 and 2024YFA1409200, National Natural Science Foundation of China with Grant Nos.~12225407, 12434005, 12074174, 12404173, 12322402 and 12274206, the Innovation Program for Quantum Science and Technology with Grant No.~2021ZD0302800, Postdoctoral Fellowship Program of CPSF with Grant No.~BX20240161, China Postdoctoral Science Foundation with
Grant No.~2024M751367, Jiangsu Province Excellent Postdoctoral Program with Grant No.~2024ZB021, and the Xiaomi Foundation. 
\end{acknowledgments}

\section{APPENDIX}

\subsection{Longitudinal Resistivity and Phase Diagram with $B \parallel c$}

\begin{figure}[h]
    \centerline{\includegraphics[width=9.cm]{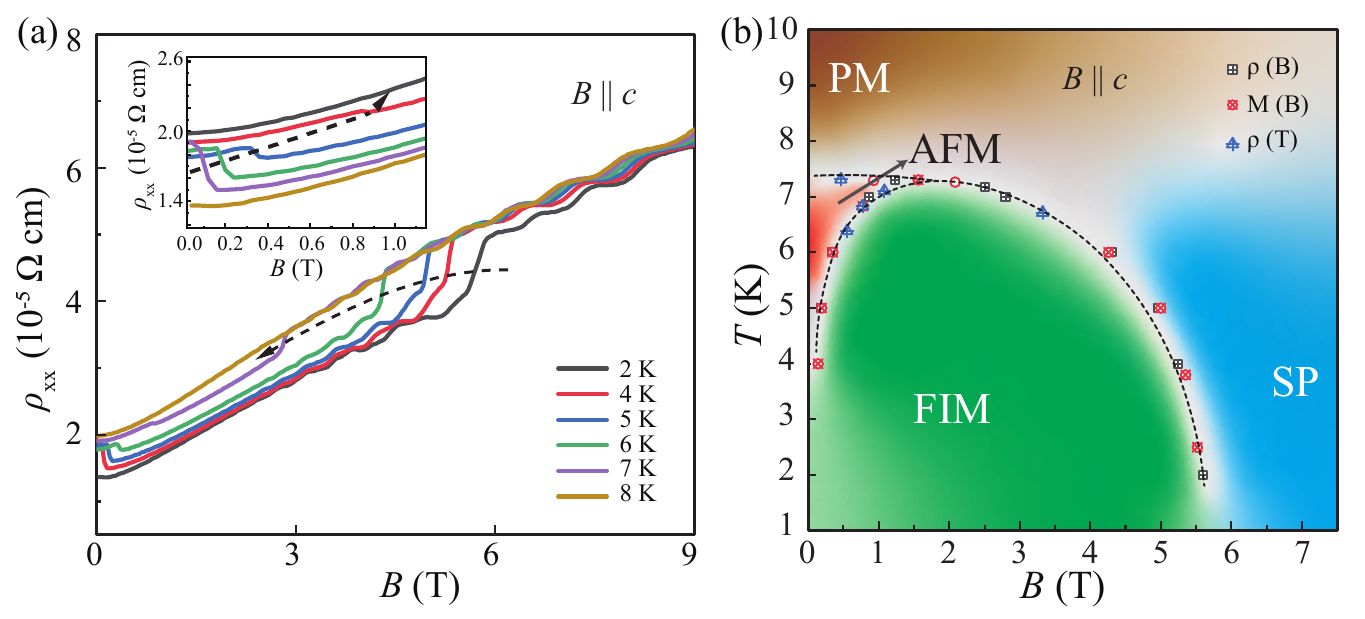}}
    \caption{\label{fig5}
    (a) Longitudinal resistivity ($\rho_{xx}$) as a function of magnetic field ($B$) along the magnetic easy axis ($B \parallel c$). Dashed arrows illustrate the evolution of the critical fields. (b) Temperature-magnetic field phase diagram of NdAlSi for $B \parallel c$. Dotted lines are guides to the eye for the phase boundaries.
    }
\end{figure}

Fig.~\ref{fig5}(a) presents the longitudinal isothermal resistivity with the magnetic field applied along the $c$-axis. At temperatures above $T_{2}$, the magnetoresistance exhibits a metallic behavior. For temperatures in the range $T_{1} < T < T_{2}$, there is an abrupt decrease in the resistivity at intermediate fields, which can be attributed to the reconstruction of the Fermi surface resulting from different spin configurations. This behavior distinguishes three phases: antiferromagnetic (AFM) at low fields, ferrimagnetic (FIM) at intermediate fields, and spin-polarized (SP) at high fields. For temperatures below $T_{1}$, the initial drop at low fields disappears, but the sharp increase at high fields remains. The phase diagram with the magnetic field along the $c$-axis, shown in Fig.~\ref{fig5}(b), is constructed based on these resistivity and magnetization data.

\subsection{Quantum Oscillations}

\begin{figure}[h]
    \centerline{\includegraphics[width=7.cm]{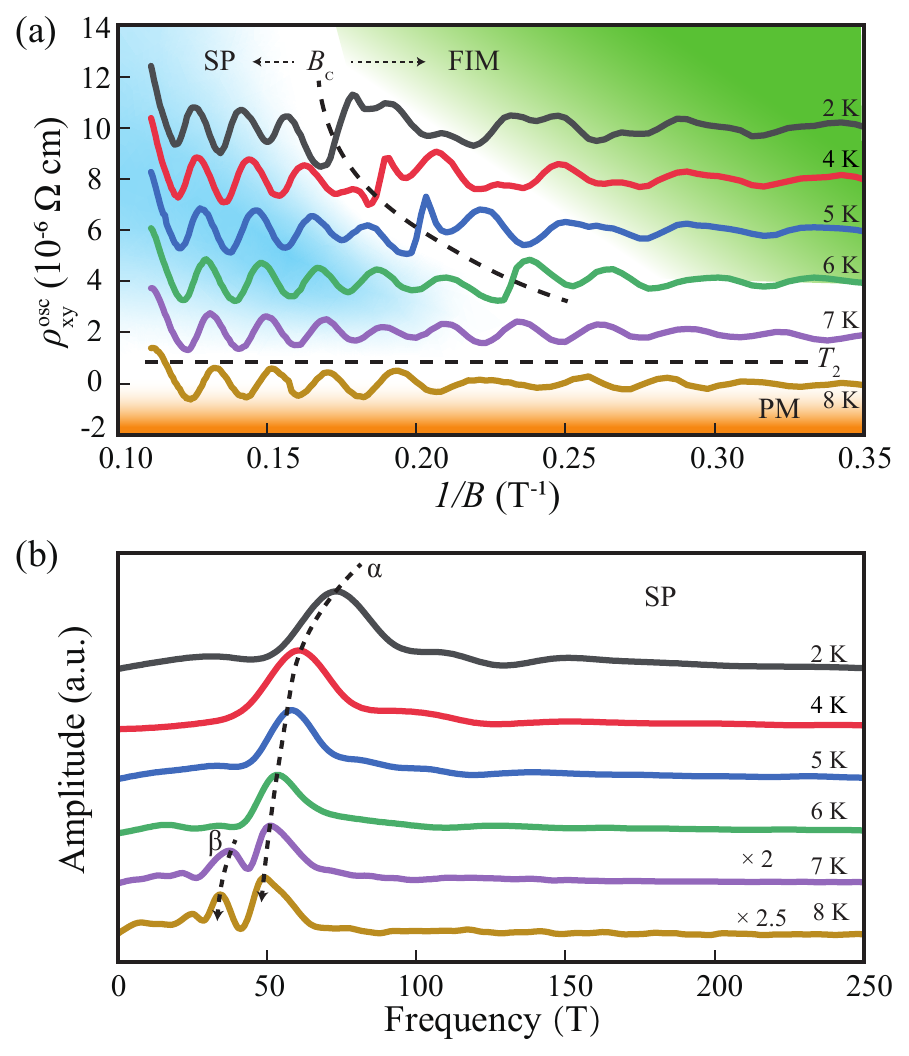}}
    \caption{\label{fig6}
    (a) The oscillatory Hall resistivity $\rho_{xy}^{\rm osc}$ as a function of inverse field. The curves are vertically offset by $2 \times 10^{-6}~\Omega~\text{cm}$ for clarity. The horizontal dashed line marks the temperature $T_{2}$, above which the system is in the paramagnetic state (PM). The dashed curve denotes the transition field $B_c$ between the spin-polarized (SP) and ferrimagnetic (FIM) states. (b) The fast Fourier transform spectra of the oscillations for the SP state. Arrows are guides to the eye for the temperature dependence of the frequency.
    }
\end{figure}

The Hall resistivity displays oscillatory behavior, particularly at fields above 3~T. Fig.~\ref{fig6}(a) shows $\rho_{xy}^{\rm osc}$ after subtracting a smooth background, plotted as a function of the inverse magnetic field ($1/B$). The plot is divided into three phases---paramagnetic (PM), SP, and FIM---separated by dashed lines.

To extract the quantum oscillation frequencies, a Fourier analysis was performed for high fields, focusing on the SP state. Fig.~\ref{fig6}(b) reveals a peak corresponding to the characteristic frequency $\alpha$ below $T_{2}$. Another characteristic frequency $\beta$, which is smaller than $\alpha$, emerges in the PM state. Notably, the frequency $\alpha$ exhibits temperature dependence, decreasing as the temperature increases. While oscillation frequencies typically hardly exhibit temperature dependence, the observed temperature-dependent frequency in NdAlSi is unusual. This temperature dependence is consistent with observations in $\rho_{xx}^{\rm osc}$ as reported in Refs.~\onlinecite{gaudet2021weyl, PhysRevResearch.5.L022013}. This behavior can be attributed to exchange interactions between Weyl fermions and the rare-earth 4$\textit{f}$ moments, which persist into the SP state due to finite 4$\textit{f}$ polarization. These interactions can influence the Weyl nodes and Fermi surfaces, contributing to phenomena such as negative magnetoresistance and the anomalous Hall effect.

\subsection{Negative Magnetoresistance with Current Parallel to the [110] Direction}

\begin{figure}[h]
    \centerline{\includegraphics[width=9.cm]{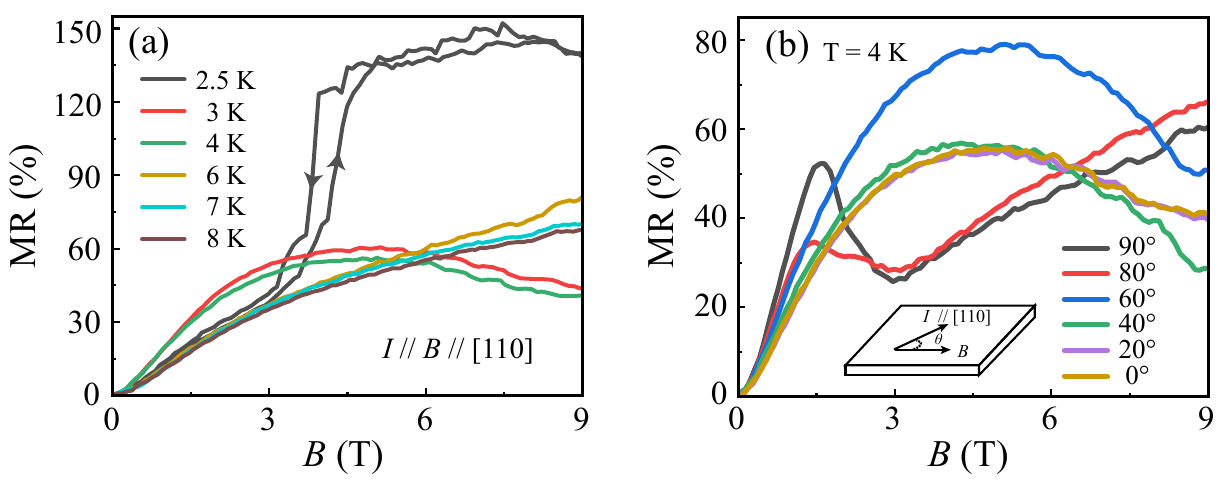}}
    \caption{\label{fig7}
     Magnetoresistance and its temperature and angle dependence with current parallel to nesting vector [110]. Panel (a) presents the temperature dependence of the magnetoresistance. Panel (b) illustrates the angle dependence of the magnetic resistance at $T= 4$ K. The inset in (b) is a schematic plot of the measurement setup.
	}
\end{figure}

Previous neutron diffraction studies reveal that the AFM Bragg peaks can be characterized by an incommensurate ordering wavevector $\mathbf{k} = (2/3+\delta , 2/3+\delta ,0)$, which also serves as the nesting vector connecting two nontrivial Fermi pockets \cite{gaudet2021weyl}. To investigate the anisotropy of magnetoresistance, measurements were performed with the current parallel to the nesting vector [110], as shown in Fig.~\ref{fig7}. 

The magnetoresistance at different temperatures in Fig.~\ref{fig7}(a) reveals three distinct behaviors. At $T = 2.5$~K, the magnetoresistance exhibits a similar trend with the behavior shown in Fig.~\ref{fig2}(a). At $T = 4$~K and $T = 5$~K, the resistance reaches its maximum around 4~T, resembling the behavior observed in Fig.~\ref{fig2}(b). For $T \geq 7$~K, no sign of saturation is observed within the measured field range.
The magnetoresistance with different field directions at 4~K is presented in Fig.~\ref{fig7}(b). When $\theta \simeq 90^\circ$ ($i.e.$, $B \perp I$), an abrupt jump is observed, likely caused by magnetic domain effects\cite{suzuki2019singular}. Additionally, the negative magnetoresistance is most pronounced when $\theta \simeq 60^\circ$ ($i.e.$, $B \parallel a$).

\subsection{Angle Dependence of the Magnetoresistance}

\begin{figure}[ht]
    \centerline{\includegraphics[width=4.5cm]{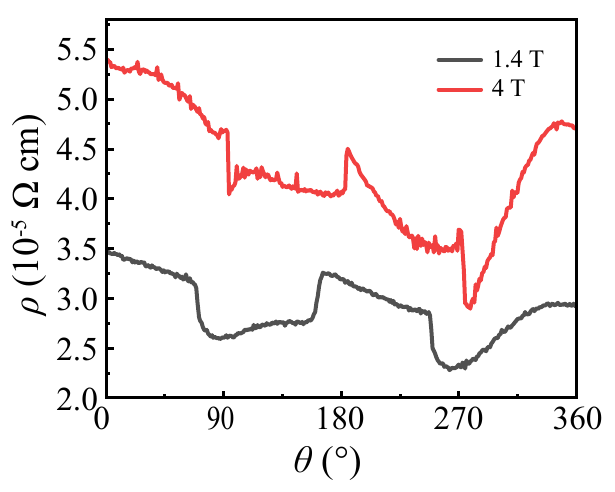}}
    \caption{\label{fig8}
    Angle dependence of magnetoresistance at $T= 4$ K under 1.4~T and 4~T. The measurement configuration is shown in the inset of Fig. \ref{fig7}(b).}
\end{figure}

Figure~\ref{fig8} presents the angle dependence of magnetoresistance at 4~K as the magnetic field is varied across different angles within the $a$-$b$ plane. At each field, distinct jumps in resistivity are observed at specific angles, which may be attributed to transitions between different magnetic domain configurations.


%

\end{document}